# Preliminary Assessment of hands motor imagery in theta- and beta-bands for Brain-Machine-Interfaces using functional connectivity analysis

J. A. Gaxiola-Tirado*, Eduardo Iáñez, Mario Ortiz, D. Gutiérrez, and José M. Azorín

*Abstract—* The use of time- and frequency-based features has proven effective in the process of classifying mental tasks in Brain Computer Interfaces (BCIs). Still, most of those methods provide little insight about the underlying brain activity and functions. Thus, a better understanding of the mechanisms and dynamics of brain activity, is necessary in order to obtain useful and informative features for BCIs. In the present study, the objective is to investigate the differences in functional connectivity of two motor imagery tasks, through a partial directed coherence (PDC) analysis, which is a frequency-domain metric that provides information about directionality in the interaction between signals recorded at different channels. Four healthy subjects participated in this study, two mental tasks were evaluated: Imagination of the movement of the right hand or left hand. We carry out the differentiation of these tasks through two different approaches: on one hand, the traditional one based on spectral power; on the other hand, an approach based on PDC. The results showed that EEG-based PDC analysis provides additional information and it can potentially improve the feature selection mainly in the beta frequency band.

**Keywords**: PDC, brain connectivity, motor imagery, hand MI, BCI

## I. INTRODUCTION

A brain-computer interface (BCI) is a communication system that allows the user to communicate with the external world through their brain's activity without the assistance of peripheral nerves and muscles [1]. The generation of successful control commands in these systems is achieved through five consecutive steps: signal acquisition, preprocessing, feature extraction, classification and the association with a control command [2]. The key problem in the analysis of EEG signals is the feature extraction process. In this framework, various techniques such as time-domain analysis, power spectral estimation, and wavelet transform have been investigated for channel selection in the processing of EEG signals.

Most training methods for non-invasive electroencephalogram (EEG)-based BCI systems involve performing a particular cognitive task, such as motor imagery (MI). In this context, classification of different mental tasks such as left/right hand MI, left-right leg IM or cognitive tasks have been investigated [3]. MI causes distinctive patterns in the electrical activity of the sensory-motor cortex, mainly in alpha (8–12 Hz) and beta (13–30 Hz) frequency bands. Then, mental tasks differentiation has been employed to control systems such as wheelchairs or a planar robot [4].

In this paper, we present the differentiation of left/right hand MI based on traditional approach based in EEG power. Then, we introduce a complementary study of the brain connectivity presented in these two tasks. Thus, a better understanding of the mechanisms and dynamics of brain activity is useful in order to obtain informative features for BCI. Therefore, we use Partial Directed Coherence (PDC) analysis, which is a frequency domain metric that provides information about directionality in the interaction between EEG signals recorded at different channels.

## II. METHODS

### A. Partial Directed Coherence

The partial directed coherence (PDC) is a frequency domain measure of the relationships (information about directionality in the interaction) between pairs of signals in a multivariate data set for application in functional connectivity inference in neuroscience [5]. If one assumes a set $S = \{x_m(n), 1 \leq m \leq M\}$ of M EEG signals (simultaneously observed time series), is adequately represented by a multivariate autoregressive (MVAR) model of order *p*, or simply MVAR(*p*):

$$\mathbf{x}(n) = \sum_{k=1}^{p} \mathbf{A}_k \mathbf{x}(n-k) + \mathbf{e}(n) \tag{2}$$

Where $\mathbf{A}_1, \mathbf{A}_2, \ldots, \mathbf{A}_p$ are the coefficient matrices (dimensions M×M), containing the coefficients $a_{ij}(k)$ which represent the linear interaction effect of $x_j(n-k)$ onto $x_i(n)$ and where $\mathbf{e}(n) = [e_1(n), e_2(n), \ldots, e_M(n)]^T$, is the noise vector (uncorrelated error process). A measure of the direct causal relations (directional connectivity) of $x_j$ to $x_i$ is given by the PDC defined by [5]

$$\pi_{i \leftarrow j} = \frac{A_{ij}(f)}{\sqrt{a_j(f)a_j^T(f)}}, \qquad (3)$$

where $A_{ij}(f)$ and $\mathbf{a}_j$ are, respectively, the $i,j$ element and the $j$-th column of

$$\mathbf{A}(f) = \mathbf{I} - \sum_{k=1}^{p} \mathbf{A}_k e^{-2i\pi f k}. \qquad (4)$$

PDC values range between 0 and 1; $\pi_{i \leftarrow j}$ measures the outflow of information from channel $x_j$ to $x_i$ in relation to the total outflow of information from $x_j$ to all of the channels.

### B. EEG acquisition

The brain activity was recorded using an EEG array of 16 electrodes (gUSBAmp from g.tec) placed on the scalp following the International 10/20 System (FC5, FC1, FC2, FC6, C3, CZ, C4, CP5, CP1, CP2, CP6, P3, PZ, P4, PO3 and PO4) at a sampling frequency of 1200 Hz.

### C. Experimental procedure

Four healthy subjects (labeled as *S1*, *S2*, *S3*, and *S4*) participated in this study, all men, with ages between the 22 and 26 years old. Subjects were sitting in a comfortable chair in front of a screen that provided instructions while their EEG signals were being recorded. Visual cues indicated which of the following two mental tasks they should have performed: Imagination of the movement of the right hand (denoted as *Class 1*) or Imagination of the movement of the left hand (*Class 2*). The task the user should have thought was displayed for 2 seconds. Each one of the two tasks had a related image that identified them. Finally, the user had to think the specific task for a period of 8-10 seconds. This process was repeated 5 times per each two tasks.

### D. Preprocessing

The methods presented in this paper were implemented in the Matlab package ARfit. Signals were processed in 1s epochs (30 epochs for each subject and Class). A digital band-pass filter between 5 and 50 Hz, a notch filter with 50 Hz cut-off frequency, were applied to the data.

## III. TRADITIONAL APPROACH BASED ON SPECTRAL POWER

Signals were processed in 1s epochs (30 epochs for each subject). For each epoch, we estimated the Power Spectral Density (PSD) in the range from 8 to 30Hz using Burg's method (function pburg in Matlab) with an autoregressive model. In order to determine the model order, we utilized the reflection coefficients (function arburg in Matlab). Thus, we used the matlab function arburg with the order set to 20 to obtain the reflection coefficients. Then, the reflection coefficients were plotted, we observed that the reflection coefficients decayed to zero after order 12. This indicated that an order 12 was the most appropriate.

Once the PSD was calculated for all frequencies, channel and class. Using all epochs of the recording session, the R-squared value was calculated for each channel and frequency as the proportion of total variance due to the power difference between the two classes [6]. A higher $r^2$ value is related to a higher discrimination between classes [7].

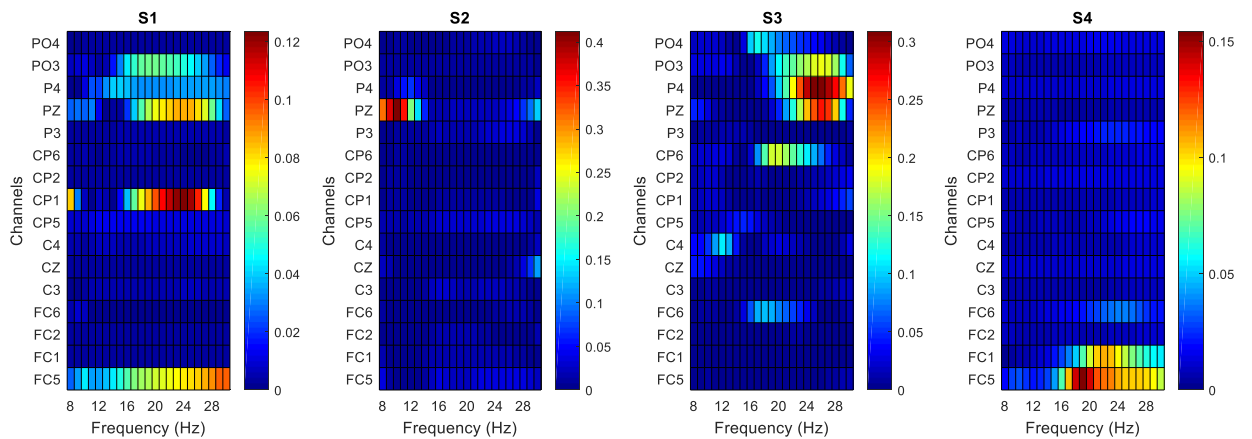

Figure 1. $r^2$ maps of *Class 1* vs *Class 2* for all subjects, channels and frequencies.

Under these conditions, the r² maps obtained for each subject are shown in Figure 1. Then, the channel and frequencies with the strongest $r^2$ (±1 Hz) were selected as optimal features. Once the features were selected, feature vectors containing PSD from the selected channels were used as input in a classifier based on Support Vector Machine (SVM) with a radial basis function as kernel (C = 512 and γ = 0.002). The parameters C and γ used in this study have been selected in function of a previous study [8]. The classifier was trained with 50% of the trials available and tested with the remaining trials. This process was repeated 100 times in a cross-validation scheme in which trials were randomly assigned to the training and testing data sets. In Table I, the channel selected, and the classification average accuracy for each subject are shown.

TABLE I: AVERAGE ACCURACY BASED ON PSD FEATURES FOR EACH SUBJECT IN TERMS OF THE PERCENTAGE OF CORRECT CLASSIFICATIONS BETWEEN CLASS 1 AND CLASS 2.

| User | Accuracy (%) | SD | Channel selected | Frequency (Hz) |
|------|--------------|------|------------------|----------------|
| S1 | 67.20 | 4.25 | CP1 | 23-25 |
| S2 | 78.50 | 2.45 | PZ | 9-11 |
| S3 | 80.30 | 3.45 | P4 | 24-26 |
| S4 | 58.00 | 5.26 | FC5 | 19-21 |

## IV. COMPLEMENTARY APPROACH BASED ON PARTIAL DIRECTED COHERENCE

### A. Complementary approach based on Partial Directed Coherence

Once preprocessing was performed, we analyzed the directed interconnections in the set of M=16 (FC5, FC1, FC2, FC6, C3, CZ, C4, CP5, CP1, CP2, CP6, P3, PZ, P4, PO3 and PO4) electrodes. In order to compute the PDC, the signals were fitted with a MVAR, where the model order was determined by the Akaike Information Criterion [9]. We analyzed the frequency range of 8 to 30 Hz. For the given set of frequencies, the PDC values from electrode $j$ to electrode $i$ ($i=1, 2,..., 16; j=1, 2,..., 16$) were obtained trough (3) for each 1s epoch (30 epochs per subject).

For each subject, we analyzed the PDC values at the frequency bands alpha (8-12 Hz) and beta (13-30 Hz). In all cases (frequency-band and direction), differences between *Class1* and *Class2* were tested using the Wilcoxon rank-sum test, under the null hypothesis that data in *Class 1* and *Class 2* have equal medians, against the alternative that they were not. The null hypothesis was rejected with p<0.001. Once differences were determined, the median over the trials for each frequency band are calculated with the purpose of comparing their magnitudes and thereby determine if the corresponding direction is predominant of *Class 1* or *Class 2*. In Figure 2 is shown the results of this process for each of them.

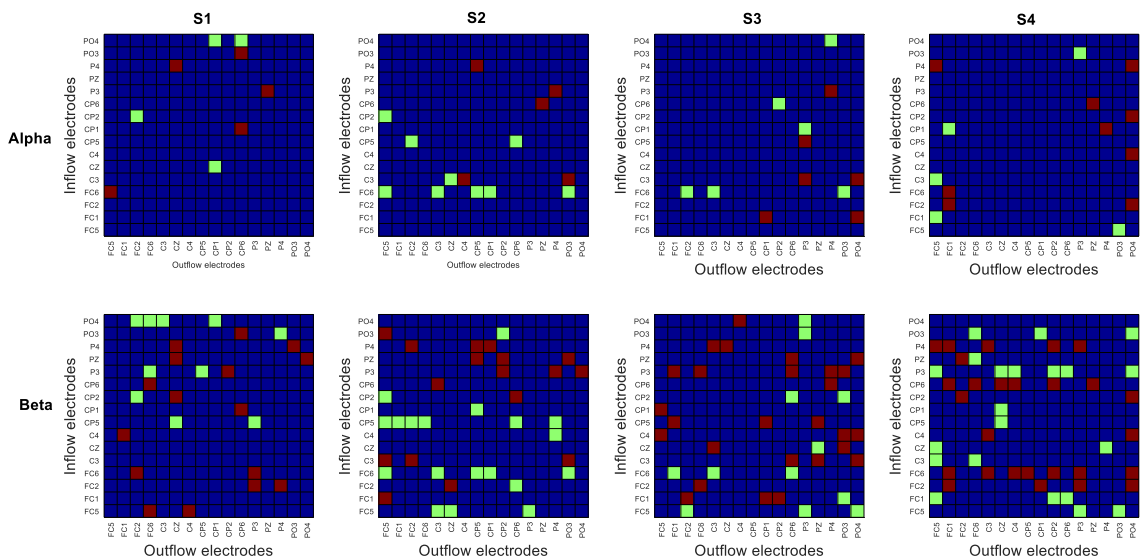

Figure 2. Interconnection directions that characterize the difference between *Class 1* and *Class 2* (p <0.001), the predominant directions of *Class 1* are shown in brown, while the predominant directions in *Class 2* are shown in green.

### B. Analysis of information flows

According to the definitions of information inflow and outflow from directed transfer function analysis [10]. From the PDC values obtained in section IV.A, the outflow information related to certain EEG channel was determined by summing-up the intensity of information flow propagated (the square of all PDC values) to other EEG channels. Meanwhile, the inflow information related to certain EEG was determined by summing-up the intensity of information flow received from other EEG channels.

The distribution of outflow and inflow related to each electrode per subject, at beta frequency band is shown in Figure 3. It can be observed the flows are different for each subject, which allows to obtain personalized information about the brain activity when performing both mental tasks.

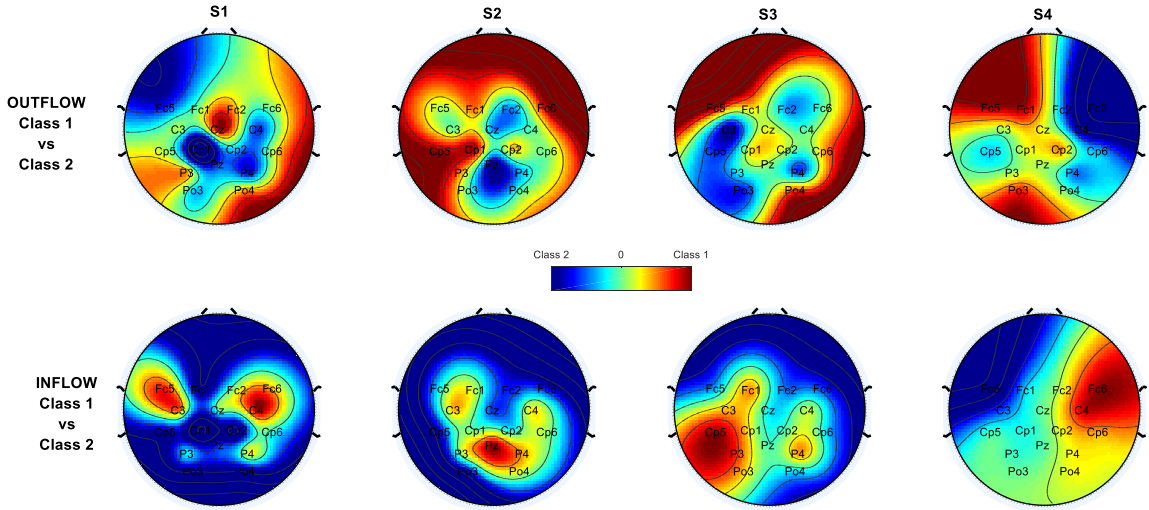

Figure 3. Outflow and inflows maps obtained through PDC when comparing Class1 and Class2.

## V. DISCUSSION

From Table 1 in section III, it can be observed that for three subjects (S1, S3 and S4) the frequencies selected are within beta frequency band. Nevertheless, this approach is based on individual electrodes, which gives poor information about the brain dynamic. So when analyzing the brain connectivity using the PDC in section IV.A. , it can be observed that during the beta band there is a greater number of statistically significant directions than in alpha band, particularly for *Class 1*, which indicates that the brain connectivity is characterized in a better way in the beta band, for which any of these significant directions could be used in the process of feature selection for the classification process.

With respect to information flows obtained through the PDC, this way of visualizing brain connectivity allows to define regions of interaction that it is not possible to define through traditional approaches based on EEG power. For instance, for subject S1, it can be observed that the outflow is predominant for channel Cz for *Class 1*, while for *Class 2*, the predominant channels are CP1, P4, Pz. Concerning to the inflows, the predominant electrodes for *Class 1* are C4, FC5 and CZ, which indicates the flow of information goes from Cz to these electrodes. On the other hand, it can be observed that for the subject S4, who obtained the lowest classification percentage (58%) the inflow and outflow from Figure 3 are more dispersed in comparison with those obtained for the subjects S1, S2 and S3. Thus, these relations of inflows and outflows can be useful in order to obtain informative features for BCI.

In conclusion, in this preliminary study was demonstrated that EEG-based PDC analysis is able to detect differences in functional connectivity presented in two motor imagery tasks. Furthermore, this analysis provides additional information and could potentially improve the feature selection process. Our future work will include a more rigorous assessment of our EEG-based connectivity analysis and extending our approach to the study of resting-state brain networks.